\documentclass[conference]{IEEEtran}
\IEEEoverridecommandlockouts
\usepackage{tikz}

\usepackage{amsmath,amssymb,amsfonts}
\usepackage{bm}
\usepackage{graphicx}
\usepackage{textcomp}
\usepackage{siunitx}
\usepackage{xcolor}
\usepackage{comment}
\usepackage{tikz}
\usepackage{algorithm}
\usepackage{algorithmic}
\usepackage{algorithmic}
\usepackage{textcomp}
\usepackage{array}       % For column alignment (m{...})
\usepackage[caption=false]{subfig}
\usepackage{tikz}
\usepackage{booktabs}    % For professional table rules
\usepackage{array}       % For column alignment (m{...})
\usepackage{color}
\usepackage{tikz}
\usepackage[normalem]{ulem}
\useunder{\uline}{\ul}{}
\usepackage{pgfplots}
\usepgfplotslibrary{groupplots}
\usepackage{times}
\usepackage{algorithmic}
\usepackage{amssymb}

\usepackage{amsxtra}
\usepackage{multirow}
\usepackage{mathtools}
\usepackage{algorithm}
\usepackage[normalem]{ulem}
\usepackage{tabu}

\usepackage{comment}
\usepackage{makecell} % Add this in your preamble if not already included
\usepackage{hyperref}
\hypersetup{hidelinks=true}
\def\BibTeX{{\rm B\kern-.05em{\sc i\kern-.025em b}\kern-.08em
    T\kern-.1667em\lower.7ex\hbox{E}\kern-.125emX}}
\AtBeginDocument{\definecolor{tmlcncolor}{cmyk}{0.93,0.59,0.15,0.02}\definecolor{NavyBlue}{RGB}{0,86,125}}
\usepackage{balance}
\usepackage{hyperref}
\usepackage{url}
\usepackage{faktor}

\usepackage{graphicx}

\usepackage{cite}
\usepackage{amsmath,amssymb,amsfonts}
\usepackage{graphicx,color}
\usepackage{textcomp}
\usepackage{xcolor}
\usepackage{hyperref}
\hypersetup{hidelinks=true}
\usepackage{algorithm,algorithmic}

\def\BibTeX{{\rm B\kern-.05em{\sc i\kern-.025em b}\kern-.08em
    T\kern-.1667em\lower.7ex\hbox{E}\kern-.125emX}}

\DeclareMathAccent{\mathH}{\mathalpha}{operators}{"7D}

\definecolor{green}{RGB}{0,128,0}
\definecolor{orange}{RGB}{255,165,0}
\definecolor{dimgray85}{RGB}{85,85,85}
\definecolor{gainsboro229}{RGB}{229,229,229}
\definecolor{lightgray204}{RGB}{204,204,204}
\definecolor{darkgray176}{RGB}{176,176,176}
\definecolor{gray128}{RGB}{100,100,100}    % Mid gray

\def\BibTeX{{\rm B\kern-.05em{\sc i\kern-.025em b}\kern-.08em
T\kern-.1667em\lower.7ex\hbox{E}\kern-.125emX}}
\usepackage[acronym]{glossaries}
\makeglossaries
\newacronym{iot}{IoT}{Internet of Things}
\newacronym{rf}{RF}{Radio Frequency}
\newacronym{owc}{OWC}{Optical Wireless Communication}
\newacronym{gnn}{GNN}{Graph Neural Network}
\newacronym{minlp}{MINLP}{Mixed-Integer Nonlinear Programming}
\newacronym{ofdm}{OFDM}{Orthogonal Frequency Division Multiplexing}
\newacronym{BP}{BP}{Bistatic Positioning}
\newacronym{MS}{MS}{Monostatic Sensing}
\newacronym{PT}{PT}{Passive Target}
\newacronym{UE}{UE}{User Equipment}

\begin{document}

\title{Joint Mobile User Positioning and Passive Target Sensing using Optimized Sequential Beamforming}

\author{
\IEEEauthorblockN{Aymen Hamrouni\IEEEauthorrefmark{1},
Sofie Pollin\IEEEauthorrefmark{1}\IEEEauthorrefmark{2},
Hazem Sallouha\IEEEauthorrefmark{1}}

\IEEEauthorblockA{\IEEEauthorrefmark{1}
\small{WaveCoRE, Department of Electrical Engineering (ESAT), KU Leuven, Leuven, Belgium}\\
\IEEEauthorblockA{\IEEEauthorrefmark{2}
\small{Interuniversity Microelectronics Centre (IMEC), Leuven, Belgium}}
\small{Email: \{aymen.hamrouni, sofie.pollin, hazem.sallouha\}@kuleuven.be}}
}

\maketitle

\begin{abstract}
Integrated sensing and communication (ISAC) relies on monostatic sensing (MS) and bistatic positioning (BP) to enable comprehensive environmental awareness and user localization. However, existing frameworks predominantly assume static geometries and optimize these modalities independently, neglecting user mobility and sequential information sharing. In this paper, we propose a velocity-aware sequential beamforming framework that dynamically couples MS and BP in time. We derive the Cram\'er-Rao bounds (CRBs) in the position domain to formulate a non-convex resource allocation problem. Instead of relying on static weighted-sum tradeoffs, we introduce a sequential Bayesian optimization strategy where MS is executed first to construct a reliable structural prior on the UE and passive targets (PTs). This covariance prior is subsequently passed to the UE to regularize the BP estimation stage. We demonstrate that optimizing a single shared beamformer globally across both phases yields superior synergistic gains compared to a two-stage greedy approach. Simulation results validate that the shared sequential design efficiently balances limited symbol resources, achieving centimeter-level positioning accuracy for both the UE and PTs, robust velocity estimation, and a significantly reduced computational runtime.
\end{abstract}


%%%%%%%%%%%%%%%%%%%%%%%%%%%%%%%%%%%%%%%%%%%%%%%%%%%%%%%%%%%%
\section{Introduction}

Integrated sensing and communication (ISAC) has emerged as a fundamental design paradigm for future wireless networks, unifying communication, localization, and environmental awareness within a single infrastructure~\cite{11153070}. By exploiting shared spectrum and hardware, ISAC enables multi-antenna systems to simultaneously support data transmission and sensing functionalities~\cite{10982381}. Within this framework, sensing functionalities can be broadly categorized into \gls{MS} and \gls{BP}, two complementary yet intrinsically coupled modalities that are both essential for comprehensive situational awareness~\cite{10440181,10978221}. \gls{MS} utilizes the Base Station's (BS) co-located transmit and receive arrays to analyze round-trip echoes for environmental mapping and detection of \glspl{PT}~\cite{10440181}.  In contrast, \gls{BP} leverages the spatially separated link between the BS and the \gls{UE}, where the latter processes downlink signals to estimate its own position and velocity relative to the infrastructure~\cite{10978221}.

\begin{figure}[t]
    \centering
    \vspace{0.3cm}
  \fbox{\includegraphics[width=0.98\linewidth]{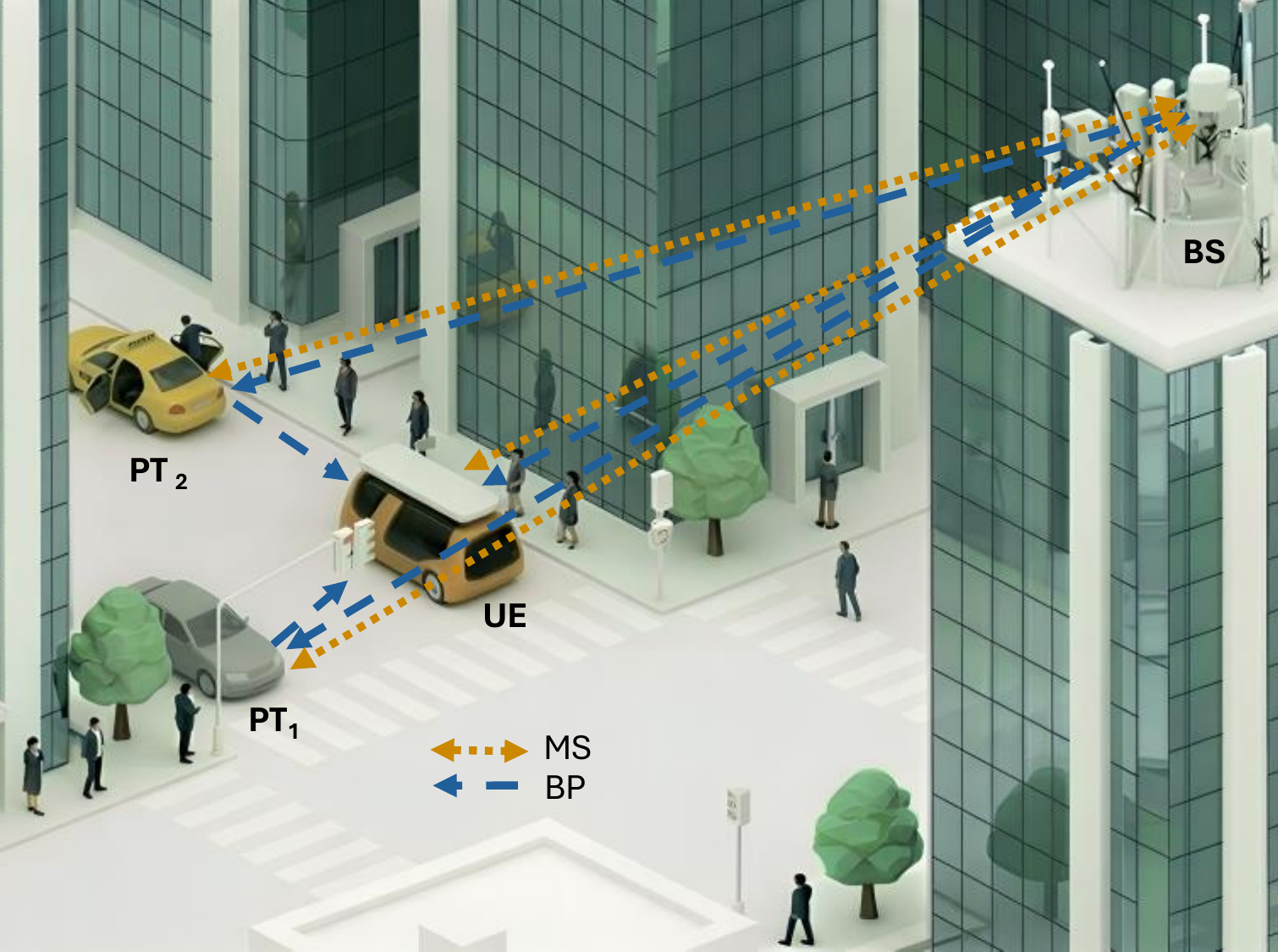}}
    \caption{Illustration of \gls{MS} and \gls{BP} in an urban canyon scenario. The BS simultaneously performs i) \gls{MS} (orange dotted lines) to detect \gls{UE} and \glspl{PT} (e.g., stationary cars) and ii) \gls{BP} (blue dashed lines) to enable localization and velocity estimation for the mobile \gls{UE}.}
    \label{fig:isac_urban_scenario}
\end{figure}
When \gls{MS}  and \gls{BP} operations share a common multi-antenna transmitter infrastructure, they fundamentally serve distinct operational requirements that compete for the same resources. From the \gls{UE} perspective, high-precision positioning and velocity estimation favor bistatic measurements that leverage the geometric diversity between spatially separated transmit and receive platforms to enhance the observability of the \gls{UE}'s spatial-temporal state.
Conversely, from the environment sensing perspective, PT detection and tracking favor monostatic radar operations, where the co-located transmitter and receiver enable coherent round-trip measurements. Consider an autonomous vehicle navigating a dense urban canyon. In such GPS-denied environments, bistatic measurements provide a crucial absolute, drift-free position and velocity reference \cite{10565792}. These RF-based updates are essential to bound the cumulative drift inherent to onboard inertial measurement units (IMUs) within sensor fusion algorithms\cite{10938202}. By anchoring the relative sensors, the vehicle maintains the stringent Position Error Bounds (PEB) and Velocity Error Bounds (VEB) required for safe, autonomous trajectory planning. Simultaneously, the serving Base Station utilizes \gls{MS} to scan the surrounding environment to detect \glspl{PT} including static objects and dynamic obstacles like pedestrians or legacy vehicles.

Fundamental spatial and time-frequency tradeoffs have been thoroughly analyzed for conventional monostatic configurations~\cite{10980163}. However, to overcome the limited coverage, severe path-loss scaling, and self-interference geometric constraints, the research community has increasingly shifted toward bistatic and multi-static deployments. For instance, Bauhofer et al.~\cite{10317749} proposed fusion techniques for processing information from multiple spatially separated sensing nodes to enable robust multi-target localization. Similarly, cooperative ISAC frameworks utilizing multiple base stations have demonstrated significant enhancements in PT localization~\cite{10681840}. Advanced multi-static architectures have been developed to improve target localization accuracy under both Line-of-Sight (LoS) and Non-Line-of-Sight (NLoS) conditions~\cite{10978602}.

Despite these advancements in spatial observability, %there remains a critical gap in understanding the fundamental resource allocation tradeoffs when \gls{MS} and \gls{BP} are jointly executed from a shared infrastructure.
only recently have attempts been made to characterize these joint tradeoffs when \gls{MS} and \gls{BP} are jointly executed from a shared infrastructure. In these recent attempts, Cram\'{e}r-Rao bound (CRB) plays a central role.
%using information-theoretic performance metrics, with the Cram\'{e}r-Rao bound (CRB) playing a central role. 
%As a fundamental lower bound on the estimation error variance, the CRB enables 
The CRB enables a principled quantification of sensing and positioning accuracy in terms of transmit covariance, capturing the spatial power allocation and beamforming strategy, and array geometry, which determines the system’s spatial resolution and parameter identifiability. Among the few works addressing this, Ge et al.~\cite{10437391} explored fusion at the filtering stage, proposing an extended Kalman-Poisson multi-Bernoulli sequential filter to integrate maps from both monostatic and bistatic sensing modalities in 5G mmWave scenarios. Zhang et al.~\cite{11029493} recently investigated the performance tradeoff between \gls{BP} and \gls{MS} within a MIMO-OFDM system, proposing a multi-objective optimization framework based on a weighted-sum CRB to design optimal spatial beamformers.
%By expressing both sensing and positioning performance within a unified CRB framework, optimization problems can be formulated to explicitly reveal and control the energy tradeoff and beam design between these two functionalities~\cite{11029493}.
Nevertheless, existing work predominantly assumes static scenarios, neglecting the inherent mobility of practical wireless systems. In reality, \gls{UE} exhibits non-negligible velocities over the sensing interval. Ignoring this mobility not only compromises model realism but also discards the valuable information embedded in Doppler shifts, which can significantly enhance both positioning and sensing performance. Furthermore, such joint operations create a fundamental resource allocation dilemma: \gls{MS} demands radar-centric beamforming patterns that maximize the illumination of uncooperative environmental scatterers, whereas \gls{BP} requires beamforming strategies that guarantee robust signal reception at the mobile vehicle. \gls{MS} and \gls{BP} directly compete for finite symbol resources within a fixed coherence interval. % Allocating more OFDM symbols to \gls{MS} increases target-related  information but inherently reduces the bistatic observations available for the \gls{UE}, and vice-versa.
To the best of our knowledge, the joint  \gls{BP} and \gls{MS} with mobile user scenarios, along with the corresponding beamforming design problem, are open research questions.

In this paper, we propose a joint \gls{MS} and \gls{BP} sensing framework that explicitly models \gls{UE} mobility and exploits the information coupling between both sensing modalities. Unlike existing static approaches, we develop a velocity-aware, sequential beamforming design that dynamically balances positioning and environmental sensing over time. Our main contributions are summarized as follows:  
\begin{itemize}
    \item We derive the CRB for joint \gls{MS} and \gls{BP} in a practical mmWave MIMO-OFDM system under \gls{UE} mobility. By extracting delay, angle, and Doppler parameters in the channel domain, we map them to 2D position and 2D velocity to formulate a non-convex beamforming optimization under a fixed energy budget.
    \item We establish a sequential strategy that executes \gls{MS} first to construct a reliable prior on both the \gls{UE} state and environmental \glspl{PT}. This prior is then parameterized via its covariance and passed to the \gls{BP} stage. This ordering is critical since standalone \gls{BP} is poorly conditioned for \gls{PT} estimation due to limited geometric diversity and nuisance parameters (e.g., clock bias). However, injecting the \gls{MS}-derived covariance regularizes the problem, enabling the \gls{UE} to jointly sense the \glspl{PT} and refine its own position.
    \item We introduce an adaptive temporal allocation policy that optimizes the distribution of finite symbol resources between the two stages, balancing direct measurement quality with the gains of covariance-level information sharing.
\end{itemize}
Our simulation results reveal that, within a given coherent block, optimizing separate greedy beamformers for the \gls{MS} and \gls{BP} fails to maximize the joint posterior information. Instead, a single shared beamformer, optimized globally across both stages achieves superior \gls{UE} PEB, \glspl{PT} PEB, and \gls{UE} VEB. This performance gain stems from coherent coupling as shared illumination pattern simultaneously strengthens both the \gls{MS} prior and the \gls{BP} likelihood, whereas separate beamformers dilute energy through stage-wise reconfiguration. Furthermore, this proposed shared formulation requires significantly less computational runtime than the two-stage greedy approach by avoiding beam-switching, reducing the effective search space.

%%%%%%%%%%%%%%%%%%%%%%%%%%%%%%%%%%%%%%%%%%%%%%%%%%%%%%%%%%%%
\section{System Model and Problem Formulation}
\label{sec:system_model}

\subsection{System Setup}

\begin{table}[t]
\centering
\caption{CPI for \gls{UE} Constant-Velocity Assumption by Scenario}
\label{tab:cpi_bounds}
\begin{tabular}{lcccc}
\hline
\textbf{Scenario} & $f_c$ (GHz) & $v_U$ (m/s)  & $T_{\mathrm{CPI}}$ \\
\hline
Pedestrian & 28 & 5 & $\lesssim 1.0$ ms \\
Vehicle (urban) & 28 & 30 & $\lesssim 180$ µs \\
High-speed vehicle & 28 & 50 & $\lesssim 100$ µs \\
UAV (mmWave) & 60 & 50 & $\lesssim 50$ µs \\
\hline
\end{tabular}
\end{table}

We consider a MIMO–OFDM setup with $M$ subcarriers, where a BS equipped with $N_B$ co-located transmit-receive antennas sends pilot signals over $L$ slots at center frequency $f_c$ to a \gls{UE} with $N_U$ antennas. The \gls{UE} exploits these received pilots to determine its position and velocity in \gls{BP}. In \gls{MS}, the BS collects the echoes from \glspl{PT} to infer their positions and from the \gls{UE} to infer its position and velocity. In this configuration, each PT contributing to \gls{MS} corresponds to a single multipath component in \gls{BP}.

We assume the \gls{UE} has a position  $\mathbf{p}_U \in \mathbb{R}^2$  and moving with a constant velocity vector $\mathbf{v}_U \in \mathbb{R}^2$, , where $v_U = \|\mathbf{v}_U\|$ denotes its speed magnitude, during a coherent processing interval (CPI).  The $K$ \glspl{PT} reflectors are assumed to be static during CPI (e.g., roadside infrastructure, building facades, or parked vehicles) with unknown positions. Considering the BS's positioning signals across $L$ slots, we assume that each slot has a duration of $T_{\text{slot}}$ and contains $P$ OFDM pilot symbols each of duration $T_{\text{sym}}$. The transmission start time of the $(l,p)$-th pilot symbol is $
t_{l,p} = (l-1)T_{\text{slot}} + (p-1)T_{\text{sym}}$, where $l=1,\ldots,L,\; p=1,\ldots,P$.
\begin{comment}
    The CPI spans from the center of the first pilot to the center of the last pilot and is given by [addREF]
\begin{equation}
T_{\mathrm{CPI}} = (L-1)T_{\text{slot}} + P T_{\text{sym}}.
\end{equation}

\end{comment}

The constant-velocity assumption holds provided that $T_{\mathrm{CPI}} = t_{L,P}  \le T_c$, where the coherence time, $T_c$, is approximated from the maximum Doppler spread as \cite{4698577}: 
\begin{equation}
T_c \approx \frac{\lambda}{2 v_{\text{rel}}},
\end{equation}
with wavelength $\lambda = c/f_c$ and relative velocity $v_{\text{rel}} = v_U$ since the BS and \glspl{PT} are assumed static. Under this condition, the small-scale fading coefficients remain approximately constant over the CPI. Table~\ref{tab:cpi_bounds} provides quantitative CPI limits for practical scenarios based on realistic carrier frequencies and \gls{UE} velocities\cite{11226246}. The transmit signal on the $m$-th subcarrier of the $p$-th symbol in the $l$-th slot is
\begin{equation}
    \mathbf{x}_{l,p,m} = \mathbf{f}_l s_{p,m},
    \label{eq:tx_signal}
\end{equation}
where $\mathbf{f}_l \in \mathbb{C}^{N_B}$ is the beamformer used in slot $l$, and $s_{p,m}$ is a unit-modulus pilot on subcarrier $m$ of symbol $p$. For clarity, we adopt the following notation convention: variables denoted as $\acute{X}$ correspond to MS-related quantities, while variables denoted as $\mathH{X}$ correspond to BP-related quantities.

\subsection{Receive Signal for  \gls{MS} and \gls{BP}}

The signal collected for \gls{MS} at the co-located BS receiver is denoted as \cite{11029493}:
\begin{equation}
    \acute{\mathbf{y}}_{l,p,m} = \acute{\mathbf{H}}_{l,p,m} \mathbf{x}_{l,p,m} + \acute{\mathbf{z}}_{l,p,m},
    \label{eq:ms_rx}
\end{equation}
with $\acute{\mathbf{z}}_{l,p,m} \sim \mathcal{CN}(\mathbf{0}, \sigma^2 \mathbf{I}_{N_B})$ is the AWGN at the BS and $\acute{\mathbf{H}}_{l,p,m} \in \mathbb{C}^{N_B \times N_B}$ is the round-trip BS-\glspl{PT}/\gls{UE} channel at subcarrier $m$, symbol $p$, and slot $l$, expressed as:
\begin{align}
    \acute{\mathbf{H}}_{l,p,m}
    &=
    \acute{\beta}_0
    e^{-j 2\pi m \Delta f \acute{\tau}_0}
    e^{j 2\pi \acute{\nu}_0 t_{l,p}}
    \mathbf{a}_B(\theta_0)
    \mathbf{a}_B^\text{H}(\theta_0)
    \nonumber \\
    &\quad +
    \sum_{k=1}^{K}
    \acute{\beta}_k
    e^{-j 2\pi m \Delta f \acute{\tau}_k}
    \mathbf{a}_B(\theta_k)
    \mathbf{a}_B^\text{H}(\theta_k),
    \label{eq:ms_channel_corrected}
\end{align}

Here, $K$ denotes the number of \glspl{PT} and for path $k$, $\acute{\beta}_k$, $\acute{\tau}_k$, and $\theta_k$ are the complex gain, delay, and angle-of-departure (AOD), respectively. The \gls{UE} is treated as a \gls{PT} in \gls{MS} and indexed by $k=0$ and Doppler $\acute{\nu}_0$ exists only for LoS \gls{UE} path.  The noise power is $\sigma^2 = F N_0 \Delta f$, where $F$ is the noise figure, $N_0$ is the single-sided noise PSD, and $\Delta f$ is the subcarrier spacing. The vector $\mathbf{a}_B(\cdot) \in \mathbb{C}^{N_B}$ denotes the steering vectors at the BS.

The received signal at the \gls{UE} for \gls{BP} is expressed as \cite{11029493}:
\begin{equation}
      \mathH{\mathbf{y}}_{l,p,m}  = \mathbf{W}^{\text{H}} \mathH{\mathbf{H}}_{l,p,m} \mathbf{x}_{l,p,m} + \mathH{\mathbf{z}}_{l,p,m},
    \label{eq:bp_rx}
\end{equation}
where $\mathH{\mathbf{z}}_{l,p,m} \sim \mathcal{CN}(\mathbf{0}, \sigma^2 \mathbf{I}_{N_U})$ denotes AWGN at the \gls{UE} and $\mathbf{W} \in \mathbb{C}^{N_U \times N_{U_\text{RF}}}$ is the analog combining matrix at the \gls{UE}, with $N_{U_\text{RF}}$ denoting the number of RF chains. The matrix $\mathH{\mathbf{H}}_{l,p,m} \in \mathbb{C}^{N_U \times N_B}$ is the BS–\gls{UE} channel at subcarrier $m$, symbol $p$, and slot $l$, modeled as:
\begin{align}
    \mathH{\mathbf{H}}_{l,p,m}
    =
    \sum_{k=0}^{K}
    \mathH{\beta}_k
    e^{-j 2\pi m \Delta f \mathH{\tau}_k}
    e^{j 2\pi \mathH{\nu}_k t_{l,p}}
    \mathbf{a}_U(\psi_k)
    \mathbf{a}_B^\text{H}(\theta_k),
    \label{eq:bp_channel_corrected}
\end{align}
where $\psi_k$ denotes the  angle-of-arrival (AOA) at the \gls{UE} from path $k$ and Doppler term $e^{j 2\pi \mathH{\nu}_k t_{l,p}}$ accounts for the phase evolution due to relative motion from all paths $k$. Here, the LoS component is indexed by $k=0$ and the vector $\mathbf{a}_U(\cdot) \in \mathbb{C}^{N_U}$ denotes the steering vectors at the \gls{UE}.

\subsection{CRB-Based Performance Metrics}

The localization process in both \gls{MS} and \gls{BP} is modeled as a two-stage process where, in the first stage, channel-domain parameters are estimated, and in a second stage, these channel-domain parameters are mapped to position-domain parameters. For \gls{MS}, we define the channel-domain parameter vector as:
\begin{equation}
    \acute{\boldsymbol{\xi}}
    =
    \big[
    \boldsymbol{\theta}^\mathrm{T},
    \acute{\boldsymbol{\tau}}^{ \mathrm{T}},
    \acute{\boldsymbol{\nu}}^{\mathrm{T}},
    \acute{\boldsymbol{\beta}}^{ \mathrm{T}}_R,
    \acute{\boldsymbol{\beta}}^{\mathrm{T}}_I
    \big]^\mathrm{T}
    \in \mathbb{R}^{4K+5},
\end{equation}
where $\boldsymbol{\theta}= [\theta_0,\ldots,\theta_K]^\mathrm{T}$, $\acute{\boldsymbol{\tau}}= [\acute{\tau}_0,\ldots,\acute{\tau}_K]^\mathrm{T}$, 
 $\acute{\boldsymbol{\nu}}
 =
[\acute{\nu}_0, 0, \ldots, 0]^\mathrm{T}$, $\acute{\boldsymbol{\beta}}_R = [\Re\{\acute{\beta}_0\},\ldots,\Re\{\acute{\beta}_K\}]^\mathrm{T},$ $\acute{\boldsymbol{\beta}}_I = [\Im\{\acute{\beta}_0\},\ldots,\Im\{\acute{\beta}_K\}]^\mathrm{T}$.
The \gls{MS} position-domain parameters are written as:
\begin{equation}
    \acute{\boldsymbol{\eta}}
    =
    \big[
    \mathbf{p}_U^\mathrm{T},
    \mathbf{v}_U^\mathrm{T},
    \mathbf{p}_1^\mathrm{T},\ldots,\mathbf{p}_K^\mathrm{T},
    \acute{\boldsymbol{\beta}}^{\mathrm{T}}_R,
    \acute{\boldsymbol{\beta}}^{ \mathrm{T}}_I
    \big]^\mathrm{T}
    \in \mathbb{R}^{4K+6}.
\end{equation}
The $(i,j)$-th element of the \gls{MS} channel-domain Fisher Information Matrix (FIM), $\mathbf{I}_{\mathrm{Chan}}(\acute{\boldsymbol{\xi}})$, is given by \cite{9709118}:
\begin{align}
    [\mathbf{I}_{\mathrm{Chan}}(\acute{\boldsymbol{\xi}})]_{i,j}
    &=
    \frac{2}{\sigma^2}
    \sum_{l=1}^L \sum_{p=1}^P \sum_{m=1}^M
    \Re \left\{
        \frac{\partial \acute{\boldsymbol{\mu}}^{\mathrm{H}}_{l,p,m}}
        {\partial [\acute{\boldsymbol{\xi}}]_i}
        \frac{\partial \acute{\boldsymbol{\mu}}_{l,p,m}}
        {\partial [\acute{\boldsymbol{\xi}}]_j}
    \right\},
    \label{eq:fim_bp_doppler}
\end{align}
where $\acute{\boldsymbol{\mu}}_{l,p,m} = \acute{\mathbf{H}}_{l,p,m} \mathbf{x}_{l,p,m}$ is the noise-free observation in Eq.~\eqref{eq:ms_rx}. 
The \gls{MS} position-domain FIM can then be obtained as $\mathbf{I}_{\mathrm{Pos}}(\acute{\boldsymbol{\eta}})
    =
   \acute{ \mathbf{J}}^{ \mathrm{T}}
    \mathbf{I}_{\mathrm{Chan}}(\acute{\boldsymbol{\xi}})
    \acute{\mathbf{J}}$, with $\acute{\mathbf{J}}$ denoting the Jacobian matrix relating the position-domain parameters   $\acute{\boldsymbol{\eta}}$
and channel-domain parameters $\acute{\boldsymbol{\xi}}$. Hence, the lower bound on the estimation error variance  for \gls{UE} position and velocity and \glspl{PT} positions is written as follows:
\begin{equation}
     \mathrm{CRB_{\gls{MS}}}(p_{U},v_{U},p_k)=
    \operatorname{tr}\!\left(
        [\mathbf{I}_{\text{Pos}}(\acute{\boldsymbol{\eta}})^{-1}]_{1:2K+4,\,1:2K+4}
    \right),
    \label{eq:crb_ms}
\end{equation}
with $\operatorname{tr}(.)$ denoting the trace operation.  For \gls{BP}, we define the vector of channel-domain parameters as:
\begin{equation}
    \mathH{\boldsymbol{\xi}}
    =
    \big[
    \boldsymbol{\theta}^\mathrm{T},
    \boldsymbol{\psi}^\mathrm{T},
    \mathH{\boldsymbol{\tau}}^{ \mathrm{T}},
    \mathH{\boldsymbol{\nu}}^{ \mathrm{T}},
   \mathH{ \boldsymbol{\beta}}^{ \mathrm{T}}_R,
    \mathH{\boldsymbol{\beta}}^{\mathrm{T}}_I
    \big]^\mathrm{T}
    \in \mathbb{R}^{6K+6},
\end{equation}
where
    $\boldsymbol{\psi}= [\psi_0,\ldots,\psi_K]^\mathrm{T}$ and
    $\mathH{\boldsymbol{\nu}}= [\mathH{\nu}_0,\ldots,\mathH{\nu}_K]^\mathrm{T}$.  The position-domain parameters for BP are collected as:
\begin{equation}
   \mathH{ \boldsymbol{\eta}}
    =
    \big[
    \mathbf{p}_U^\mathrm{T},
    \mathbf{v}_U^\mathrm{T},
    \mathbf{p}_1^\mathrm{T},\ldots,\mathbf{p}_K^\mathrm{T},
    \Delta \varphi,
    \Delta t,
   \mathH{ \boldsymbol{\beta}}^{ \mathrm{T}}_R,
   \mathH{ \boldsymbol{\beta}}^{ \mathrm{T}}_I
    \big]^\mathrm{T}
    \in \mathbb{R}^{4K+8},
\end{equation}
The parameter $\Delta \varphi$ denotes the BS orientation in the \gls{UE}’s local frame, and $\Delta t$ is the clock bias capturing BS–\gls{UE} asynchronism. Analogously to MS,  the \gls{BP} position-domain FIM is written as:
\begin{equation}
    \mathbf{I}_{\mathrm{Pos}}(\mathH{\boldsymbol{\eta}})
    =
   \mathH{ \mathbf{J}}^{ \mathrm{T}}
    \mathbf{I}_{\mathrm{Chan}}(\mathH{\boldsymbol{\xi}})
    \mathH{\mathbf{J}},
\end{equation}
with
$[\mathH{\mathbf{J}}]_{i,j}
=
\partial [\mathH{\boldsymbol{\xi}}]_i /
\partial [\mathH{\boldsymbol{\eta}}]_j$.
The CRB for \gls{UE} parameters of interests (i.e., position and velocity) estimation in \gls{BP} is written as:
\begin{equation}
    \mathrm{CRB_{\gls{BP}}}(p_{U},v_{U})
    =
    \operatorname{tr}
    \left(
    [\mathbf{I}_{\mathrm{Pos}}(\mathH{\boldsymbol{\eta}})^{-1}]_{1:4,1:4}
    \right).
\end{equation}

\subsection{\gls{MS}-\gls{BP} Trade-off Problem Formulation}
\label{sec:wscrb}

To initially characterize the \gls{MS}-\gls{BP} tradeoff, we adopt a classical weighted-sum method for MOO problems, as in \cite{11029493}, which yields a  weak Pareto frontier between \gls{MS} and \gls{BP}. The formulation can be written as:
\begin{align}
   \text{(P1)}\text{ } \min_{\mathbf{F}} \text{ } &
    (1-\alpha)\, \mathrm{CRB_{\gls{MS}}}(p_{U},v_{U},p_k) + \alpha \, \mathrm{CRB_{\gls{BP}}}(p_{U},v_{U})
    \notag \\
    \text{s.t.} \quad &
    \operatorname{tr}(\mathbf{F}\mathbf{F}^\text{H})
    \le \frac{P_B}{M},
    \label{eq:wscrb_power}
\end{align}
where $\mathbf{F} = [\mathbf{f}_1,\ldots,\mathbf{f}_L] \in \mathbb{C}^{N_B \times L}$ is  the beamformer matrix that $\mathbf{F}$ collects the $L$ beamformers and $P_B$ is the BS transmit power budget. The constraint is normalized so that the total power across all $M$ subcarriers equals $P_B$. $\alpha \in [0,1]$ controls the objective trade-off between  \gls{MS} and \gls{BP}. The problem (P1) is non-
convex but can be relaxed to a convex Semi-Definite Programming (SDP) and can be solved efficiently using standard solvers such as CVX\cite{11029493}.

\begin{comment}
    The FDB design can be simplified to a power allocation problem over the fixed codebook $\mathbf{U}$. Let $\boldsymbol{\rho} = [\rho_1,\ldots,\rho_{2K+2}]^\mathrm{T}$ denote the non-negative power allocation vector and set
\begin{equation}
    \mathbf{V} = \mathbf{U} \operatorname{diag}(\boldsymbol{\rho}) \mathbf{U}^\text{H}.
\end{equation}
The corresponding CPA formulation is
\begin{align}
   \text{(P2)} \min_{\boldsymbol{\rho}} \quad &
    \alpha \, \text{CRB}^{''}(U)
    + (1-\alpha)\, \text{CRB}^{'}(U,T)
    \label{eq:wscrb_cpa} \\
    \text{s.t.} \quad &
    \operatorname{tr}\big(\mathbf{U} \operatorname{diag}(\boldsymbol{\rho}) \mathbf{U}^\text{H}\big)
    \le \frac{P_B}{M}, \quad
    \boldsymbol{\rho} \ge \mathbf{0},
    \label{eq:wscrb_cpa_const}
\end{align}
where $\boldsymbol{\rho} \ge \mathbf{0}$ is enforced elementwise matrix. Using the same transformations as in \cite{11029493}, the problem can be recasted as a convex optimization problem solvable with standard tools.
\end{comment}

\section{Bayesian-based Sequential \gls{MS}-\gls{BP} Design}

The previous formulation (P1) considers the beamforming design of \gls{MS} and \gls{BP}  independently through a weighted-sum objective, ignoring time resources and information sharing between the two operations. We extend the state-of-the-art MS-BP joint optimization framework to a sequential setting in which the spatial beamformers for BP and MS are jointly optimized, along with the pilot allocation ratio between \gls{MS}  and \gls{BP}. In this setting, \gls{MS} is executed first so the BS obtains prior information about the \gls{UE} and \glspl{PT}. Then, \gls{BP} is optimized to refine the \gls{UE} state and infer the \glspl{PT} positions given this prior. Both the BS and \gls{UE} share a common symbol resource. Let that be the total number of pilot symbols $\kappa_{\text{Total}}$ in the coherence time $T_c$. $\kappa_{\text{Total}}$  is partitioned into three distinct phases and is written as:
\begin{equation}
\kappa_{\text{Total}} 
= \kappa_{\text{\gls{MS}}} 
+ \kappa_{\text{\gls{MS}-\gls{BP}}} 
+ \kappa_{\text{\gls{BP}}}, 
\end{equation}
where $\kappa_{\text{\gls{MS}}}$ and $\kappa_{\text{\gls{BP}}}$ are the total active sensing/positioning symbols, and $\kappa_{\text{\gls{MS}-\gls{BP}}}$ represents the number of symbols needed for the BS to transmit the computed \gls{MS} covariance matrix to the \gls{UE} for  \gls{BP} via a communication control channel.  We define the active \gls{MS} allocation ratio $\rho_{\text{\gls{MS}}}$ as the fraction of the remaining allocable symbols dedicated to the monostatic sensing, which is given by:
\begin{equation}
\rho_{\text{\gls{MS}}} \triangleq \frac{\kappa_{\text{\gls{MS}}}}{\kappa_{\text{Total}} - \kappa_{\text{\gls{MS}-\gls{BP}}}} , \quad 0 \le \rho_{\text{\gls{MS}}} \le 1. 
\end{equation}
The parameter $\rho_{\text{\gls{MS}}}$ dictates the optimal switching point between \gls{MS} and \gls{BP}, directly controlling the performance of \gls{MS} at the BS and the amount of prior information versus the remaining resources available for the \gls{BP} posterior.

While \gls{MS} and \gls{BP} observe the same physical parameters (i.e., \gls{UE} and \glspl{PT} states), their measurements are collected over the disjoint pilot slots $\kappa_{\text{\gls{MS}}}$ and $\kappa_{\text{\gls{BP}}}$ and are affected by distinct nuisance parameters (e.g., different hardware and clock biases). Let the shared position-domain parameter vector be defined as\footnote{Note that the channel gain parameters $\acute{\boldsymbol{\beta}}_R,
    \acute{\boldsymbol{\beta}}_I, \mathH{ \boldsymbol{\beta}}_R,
   \mathH{ \boldsymbol{\beta}}_I, $ are excluded, as their contribution to the position estimation is negligible \cite{11029493}. In the 2D position and 2D velocity estimation, the \gls{BP} clock offset $\Delta_t$ and orientation offset $\Delta_{\varphi}$ are likewise marginalized.} $\boldsymbol{\eta}_p=\big[
    \mathbf{p}_U^\mathrm{T},
    \mathbf{v}_U^\mathrm{T},
\mathbf{p}_1^\mathrm{T},\ldots,\mathbf{p}_K^\mathrm{T}\big]$. By marginalizing out the stage-specific nuisances via the Schur complement\cite{5571900}, we  parameterize the FIM over the same parameter vector $\boldsymbol{\eta}_p$ and obtain the equivalent position-domain FIMs for each stage, denoted as $\acute{\mathbf{I}}_{\text{Pos}}$ and $\mathH{\mathbf{I}}_{\text{Pos}}$, for \gls{MS} and \gls{BP}, respectively.  The posterior distribution via Bayes' theorem is then expressed as:
\begin{equation}
p(\boldsymbol{\eta}_p | \acute{\mathbf{y}}, \mathH{\mathbf{y}}) \propto p(\mathH{\mathbf{y}} | \boldsymbol{\eta}_p) \underbrace{p(\boldsymbol{\eta}_p | \acute{\mathbf{y}})}_{\text{\gls{MS}-based Prior}}.
\end{equation}
Consequently, the posterior FIM for $\boldsymbol{\eta}_p $ is then additive, representing the sum of the equivalent information derived from the \gls{MS}-based prior and the \gls{BP} likelihood. The resulting \gls{BP} posterior FIM  obtained from \gls{MS} prior during $\rho_{\text{MS}}$ and from \gls{BP} phase during $1-\rho_{\text{MS}}$ is written as:
\begin{equation}
\mathH{\mathbf{I}}_{\text{seq}} (\mathbf{F}_{\text{\gls{MS}}}, \mathbf{F}_{\text{\gls{BP}}}, \rho_{\text{MS}}) = \acute{\mathbf{I}}_{\text{Pos}}(\mathbf{F}_{\text{\gls{MS}}}, \rho_{\text{MS}}) + \mathH{\mathbf{I}}_{\text{Pos}}(\mathbf{F}_{\text{\gls{BP}}}, 1-\rho_{\text{MS}}). 
\label{seqfim}
\end{equation}
Here, the covariance matrix obtained from the \gls{MS} stage, $\acute{\mathbf{I}}_{\text{Pos}}^{-1}$, is passed as a prior to the \gls{UE} during the $\kappa_{\text{MS-BP}}$ symbols.

In practical ISAC systems, ensuring reliable localization requires that estimation errors remain below predefined safety thresholds. Motivated by this, we aim to minimize the deviation error obtained in the \gls{MS} and \gls{BP} sequential phases from predefined thresholds. We define the set containing the threshold of \gls{UE} position and velocity and \glspl{PT} positions as $\mathcal{S}$. For example, the threshold of the PEB for \gls{UE} in the \gls{MS} phase is denoted as $\gamma_{p_{U}}^{\text{\gls{MS}}} \in \mathcal{S}$.

A least-squares (LS)  optimization problem can then be formulated as:
\begin{equation}
\begin{aligned}
\text{(P2)} \quad \min_{\mathbf{F}_{\text{\gls{MS}}},\, 
\mathbf{F}_{\text{\gls{BP}}},\, 
\rho_{\text{MS}}} \quad
& \mathcal{L}\big(\mathH{\mathbf{I}}_{\text{seq}}
(\mathbf{F}_{\text{\gls{MS}}}, \mathbf{F}_{\text{\gls{BP}}}, \rho_{\text{MS}})^{-1},
\mathcal{S} \big) \\
\text{s.t.} \quad
& \mathrm{tr}(\mathbf{F}_{\text{\gls{MS}}} 
\mathbf{F}_{\text{\gls{MS}}}^{\mathrm{H}}) 
\le \tfrac{P_B}{M}, \\
& \mathrm{tr}(\mathbf{F}_{\text{\gls{BP}}} 
\mathbf{F}_{\text{\gls{BP}}}^{\mathrm{H}}) 
\le \tfrac{P_B}{M}, \\
& 0 \le \rho_{\text{MS}} \le 1.
\end{aligned} 
\label{Fm-star}
\end{equation}
where $\mathcal{L} (.,.)$ is a weighted squared posterior cost taking into account the deviation of the extracted bounds in the sequential \gls{MS} and posterior \gls{BP} from their respective thresholds in $ \mathcal{S}$. Weighting in $\mathcal{L}(.,.)$ could define the task priority (e.g., if emphasis is on obtaining better \gls{BP} VEB and PEB, weights on the deviation of these terms with respect to their thresholds can be set higher). This problem tackles the tradeoff between allocating more \gls{BP} symbols to improve \gls{BP} estimation and allocating more \gls{MS} symbols to build a more informative prior, which in turn improves \gls{BP}.
\begin{table}[t]
\centering
\caption{System, Channel, and Simulation Parameters}
\label{tab:system_params}
\begin{tabular}{@{}lcc@{}}
\toprule
\textbf{Parameter} & \textbf{Symbol} & \textbf{Value} \\
\midrule
\multicolumn{3}{c}{\textit{System and Signal Configuration}} \\
BS antenna elements & $N_{\text{BS}}$ & 64 \\
\gls{UE} antenna elements & $N_{\text{UE}}$ & 16 \\
Carrier frequency & $f_c$ & \SI{28}{GHz} \\
System bandwidth & $B$ & \SI{120}{MHz} \\
Number of subcarriers & $M$ & 1024 \\
BS transmit power budget & $P_B$ & \SI{-20}{dBm} \\
Number of slots & $L$ & 16 \\
Symbols per slot & $P$ & 100 \\
Clock bias & $\Delta t$ & \SI{1}{\micro\second} \\
Relative orientation offset & $\Delta_{\varphi}$ & $110^\circ$ \\
\midrule
\multicolumn{3}{c}{\textit{Noise and Environmental Model}} \\
Noise Power Spectral Density & $N_0$ & \SI{-173}{dBm/Hz} \\
Noise figure & $F$ & \SI{10}{dB} \\
Path loss exponent & $n$ & 3.5 \\
Shadow fading & $\sigma_{\text{shadow}}$ & \SI{8}{dB} \\
\glspl{PT} RCS (\gls{MS}) & $\sigma_{\text{RCS},k}$ & \SI{10}{m^2} \\
\gls{UE} RCS (\gls{MS}) & $\sigma_{\text{RCS},0}$ & \SI{10}{m^2} \\
\midrule
\multicolumn{3}{c}{\textit{Performance Thresholds}} \\
\gls{UE} PEB for MS & $\gamma_{p_{U}}^{\text{\gls{MS}}}$ & \SI{50}{cm} \\
\gls{UE} PEB for BP & $\gamma_{p_{U}}^{\text{\gls{BP}}}$ & \SI{10}{cm} \\
\gls{UE} VEB for BP & $\gamma_{v_{U}}^{\text{\gls{MS}}}$ & \SI{10}{m/s} \\
\glspl{PT} PEB for BP & $\gamma_{p_{k}}^{\text{\gls{BP}}}$ & \SI{20}{cm} \\
\glspl{PT} PEB for MS & $\gamma_{p_{k}}^{\text{\gls{MS}}}$ & \SI{100}{cm} \\
\bottomrule
\end{tabular}
\end{table}
Solving (P2) jointly over all three variables would require a bilevel optimization: for each candidate $\mathbf{F}_{\text{\gls{MS}}}$, the inner $\mathbf{F}_{\text{\gls{BP}}}$ problem must be solved conditioned on the resulting prior, which is computationally prohibitive. We therefore consider two tractable approaches.

A greedy two-stage approximation to (P2) in which the \gls{MS} beamformer 
$\mathbf{F}_{\text{\gls{MS}}}^{\star}$  is independently from (P1), with $\alpha=0$ as follows: 
\begin{equation}
\mathbf{F}_{\text{\gls{MS}}}^{\star} = \arg\min_{\mathbf{F}} \; \mathcal{L}_{\text{\gls{MS}}}\!\left(\acute{\mathbf{I}}_{\text{Pos}}(\mathbf{F}, \rho_{\text{MS}})^{-1}, \mathcal{S}\right),
\label{eq:fms_star}
\end{equation}
where $\mathcal{L}_{\text{\gls{MS}}}(.,.)$ is the LS \gls{MS}-only cost from (P1) with respect to $\mathcal{S}$. The beamformer 
$\mathbf{F}_{\text{\gls{MS}}}^{\star}$ is then  used during the symbol allocation $\rho_{\text{\gls{MS}}}^{\star}$ to build environmental prior relative to the \gls{UE} and \glspl{PT} position thresholds. Then,  $\mathbf{F}_{\text{\gls{BP}}}^{\star}$ 
is subsequently optimized to minimize the posterior cost 
given the fixed prior $\acute{\mathbf{I}}_{\text{Pos}}
(\mathbf{F}_{\text{\gls{MS}}}^{\star}, \rho_{\text{MS}})$. In this approach, the optimal beamformers are explicitly tailored and optimized to their respective phases.  We refer to it as the \textit{separate sequential beamformer} approach.

While this beamforming optimization approach provides spatial degrees of freedom and optimizes each beamformer for its respective phase, hardware or signaling constraints may dictate a single shared beamformer across the two phases. Moreover, and as it relies on a greedy two-stage procedure where $\mathbf{F}_{\text{\gls{MS}}}^{*}$ is first optimized for the \gls{MS} cost alone in (P1), and $\mathbf{F}_{\text{\gls{BP}}}^{*}$ is subsequently optimized for the posterior given with fixed prior from the \gls{MS}, the \gls{MS} beamformer might not be the optimal beamformer to produce the best prior for downstream \gls{BP}.  It prioritizes \gls{MS} first and minimizes the \gls{MS} estimation error during  $\rho_{\text{MS}}^{*}$ rather than the quality of the prior it provides to the \gls{BP} stage, since it is optimized under $\mathcal{L}_{\text{\gls{MS}}}(.,.)$, not the end-to-end posterior cost $\mathcal{L}(.,.)$. By contrast, a single shared beamformer $ \mathbf{F}_{\text{seq}}^{*}$ used across both phases (i.e., $\mathbf{F}_{\text{\gls{MS}}}^{*} = \mathbf{F}_{\text{\gls{BP}}}^{*} = \mathbf{F}_{\text{seq}}^{*}$) could be optimized directly against the final posterior cost, allowing it to jointly balance the \gls{MS} prior quality and \gls{BP} measurement quality in a single coupled optimization. Therefore, we define the following optimization problem:
\begin{equation}
\begin{aligned}
\text{(P3)} \quad \min_{\mathbf{F}_{\text{seq}},\, 
\rho_{\text{MS}}} \quad
& \mathcal{L}\big(\mathH{{\mathbf{I}}}_{\text{seq}}
(\mathbf{F}_{\text{seq}}, \rho_{\text{MS}})^{-1},
\mathcal{S} \big) \\
\text{s.t.} \quad
& \mathrm{tr}(\mathbf{F}_{\text{seq}} 
\mathbf{F}_{\text{seq}}^{\mathrm{H}}) 
\le \tfrac{P_B}{M}, \\ & \quad
0 \le \rho_{\text{MS}} \le 1,
\end{aligned} 
\label{p3}
\end{equation}
where the optimal shared beamformer $\mathbf{F}_{\text{seq}}^{*}$ is shared across both phases in Eq.~\eqref{seqfim}. We refer to this approach as the \textit{shared sequential beamformer} approach.

\begin{figure}[t]
    \centering
    \hspace{-0.3cm}
    \subfloat[\gls{MS} (\(\alpha = 0\))]{%
\includegraphics[width=0.21\textwidth]{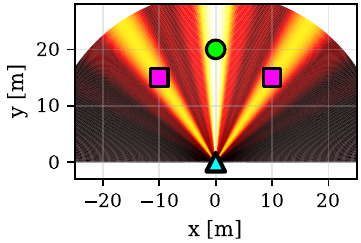}
        \label{fig:beams_ms}
    }
    \subfloat[\gls{BP} (\(\alpha = 1\))]{%
    \vspace{-0.4cm}
    \hspace{-0.3cm}
\includegraphics[width=0.27\textwidth]{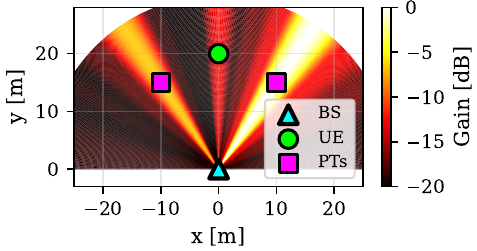}
        \label{fig:beams_bp}
    }
    \caption{Beamforming design for \gls{MS} and \gls{BP}. The \gls{UE} and \glspl{PT} are placed symmetrically at equal distances to isolate distance-dependent path-loss disparities, focusing on beamforming for MS and BP.}
    \label{fig:beam_patterns}
\end{figure}

%Although the beamformer is static across the entire coherence interval, optimizing $\rho$ ensures the system spends the correct fraction of time building the \gls{MS} prior versus refining the \gls{BP} estimate. Compared to the shared architecture, the separate architecture explicitly adapts the \gls{BP} beamformer to the \gls{MS} prior, effectively orthogonalizing information in the FIM domain and improving conditioning.

\section{Experimental Results}

In this section, we evaluate the positioning and velocity estimation performance of the sequential \gls{MS}-\gls{BP} configuration. Unless stated otherwise, all results are obtained for the parameters mentioned in  Table~\ref{tab:system_params}.  The channel-domain FIMs for \gls{BP} and \gls{MS} are computed using the Slepian-Bangs formula\cite{9709118} and mapped to the position–velocity domain via the corresponding Jacobians. The FIM of the channel parameters follows the standard diagonal assumption with no coupling between delay, angle, Doppler, and channel gain estimates across different propagation paths\cite{Wilding2018AccuracyBounds}. This approximation is justified under large array apertures and high signal bandwidth conditions, where angular and temporal resolution effectively decorrelate the parameter estimation errors.

\begin{figure}[t]
 \centering
  \hspace{-0.3cm}
 \subfloat[PEB  vs. number of \glspl{PT}]{%
\includegraphics[width=0.225\textwidth,height=0.24\textwidth]{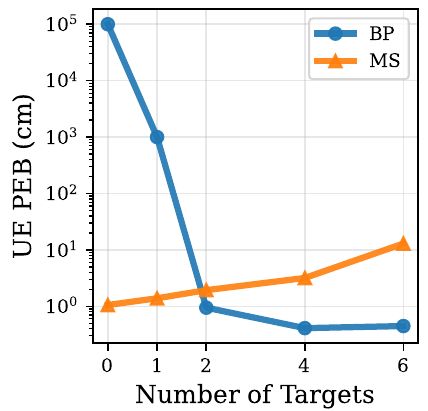}
      \label{fig:4a}}
 \qquad
 \hspace{-1cm}
 \subfloat[VEB vs. number of \glspl{PT}]{%
\includegraphics[width=0.25\textwidth,height=0.24\textwidth]{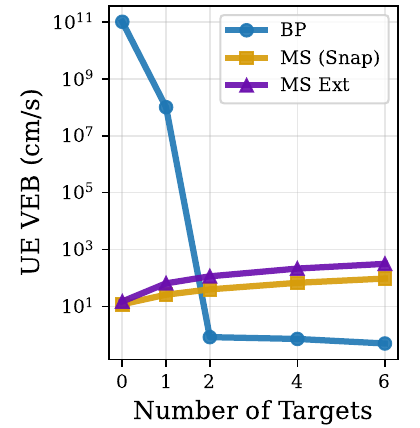}
      \label{fig:4b}}
       \caption{PEB (log-scale) vs. number of \glspl{PT} and VEB (log-scale) vs. number of \glspl{PT} for \gls{BP}, Snapshot \gls{MS}, and Extended \gls{MS}. }%
       \label{fig:4}
\end{figure}

\begin{figure}[t]
 \centering
 \vspace{0.2cm}
  \hspace{-0.8cm}
\includegraphics[width=0.245\textwidth]{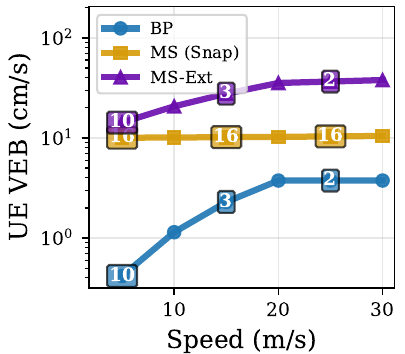}
       \caption{VEB (log-scale) vs. speed  for \gls{BP}, Snapshot \gls{MS}, and Extended \gls{MS} with $4$ \glspl{PT}. The numbers represent the number of effective slots $L_{\text{eff}} \leq L$ of observation of the \gls{UE}.}%
       \label{fig:5}
\end{figure}

\subsection{Beamforming Analysis}
\begin{figure*}[t]
 \centering
  \hspace{-0.8cm}
 \label{fig:images2}
 \subfloat[\gls{UE} PEB (log scale) ]{%
\includegraphics[width=0.25\textwidth]{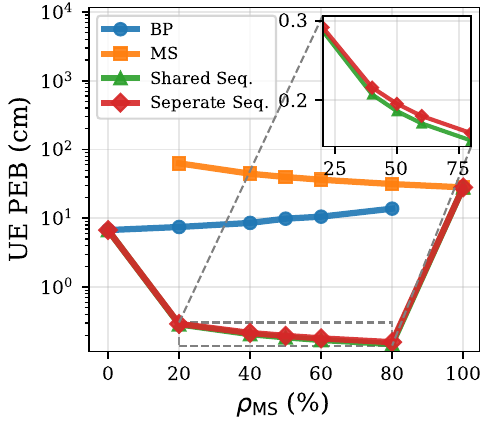}\label{fig5a}}
 \qquad
 \hspace{-0.85cm}
 \subfloat[\gls{UE} VEB (log scale)]{%
\includegraphics[width=0.25\textwidth]{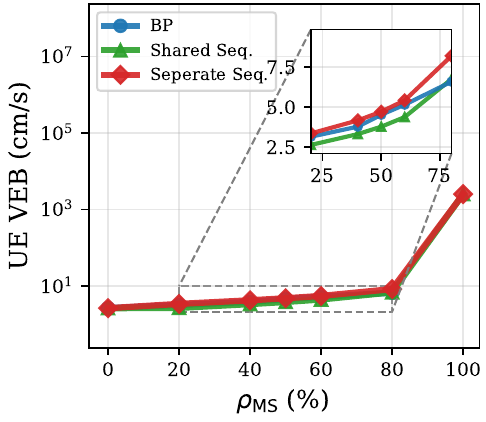} \label{fig5b}}
       \qquad
 \hspace{-0.85cm}
 \subfloat[\glspl{PT} PEB (log scale)]{%
\includegraphics[width=0.25\textwidth]{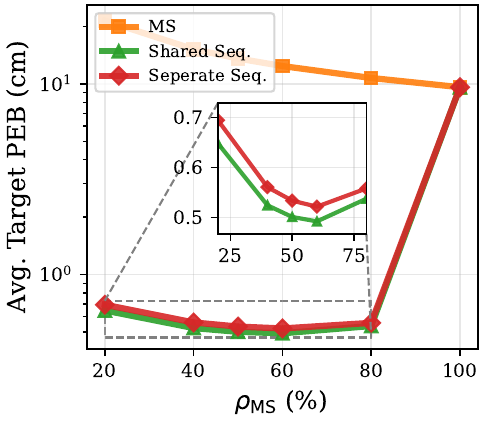}\label{fig5c}}
       \qquad
 \hspace{-0.8cm}
 \subfloat[Computational runtime (ms)]{%
      \includegraphics[width=0.268\textwidth]{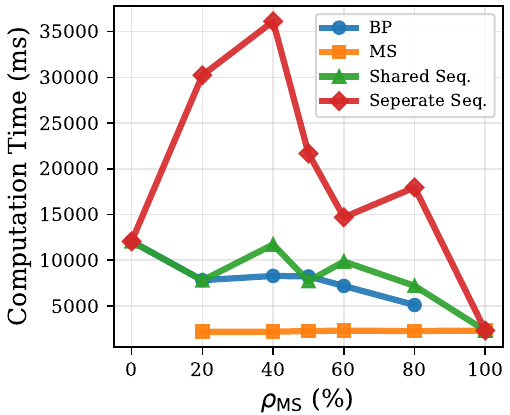}\label{fig5d}}
    \caption{Performance of full \gls{MS}, full \gls{BP}, shared sequential, and separate sequential beamforming designs vs. the \gls{MS} symbol ratio $\rho_{\text{MS}}$.}%
    \label{fig5}
\end{figure*}

\begin{comment}
\begin{figure}[b]
    \centering
    \subfloat[VEB radial for \gls{MS} and \gls{BP}]{%
        \includegraphics[width=0.24\textwidth]{figures/fig2a.pdf}
        \label{fig:2a}
    }\hspace{-0.23cm}
    \subfloat[PEB UE in \gls{MS} vs. PEB UE in \gls{BP}]{%
        \includegraphics[width=0.24\textwidth]{figures/fig2b.pdf}
        \label{fig:2b}
    }
    \caption{VEB radial and PEB UE for \gls{MS} and \gls{BP} in FDB and CPA}
    \label{fig:2}
\end{figure}
\end{comment}
To fundamentally understand the underlying spatial allocation strategies of the two MS and BP modalities, Fig.~\ref{fig:beam_patterns} visualizes the optimized spatial beam patterns for pure \gls{MS}, $\alpha = 0$, and pure \gls{BP}, $\alpha = 1$. To isolate the inherent estimation requirements of each modality from path-loss biases, the simulation geometry is deliberately constructed symmetrically. 
As observed in Fig.~\ref{fig:beams_ms}, the \gls{MS} optimization allocates spatial energy directly toward both the \glspl{PT} and the \gls{UE} with more focus on the \gls{UE}. This asymmetric power distribution in \gls{MS} is driven by the \gls{UE} velocity estimation, which relies entirely on the Doppler shift observed from the direct LoS echo.  Conversely, Fig.~\ref{fig:beams_bp} demonstrates a radically different spatial strategy for \gls{BP}. The \gls{UE} receives substantially lower LoS illumination energy, with the \glspl{PT} on the right receiving higher energy compared to the \glspl{MS} case. This occurs because FIM for \gls{BP} position and velocity estimation derives critical information not only from the LoS but also from the rich signatures of the NLoS multipath components reflected off the \glspl{PT}. Illuminating these \glspl{PT} provides the geometric diversity required to resolve the full position and velocity. Furthermore, the highest beamforming gain is directed specifically toward the rightmost PT, due to the \gls{UE} antenna array's right-tilted orientation as included in Table.~\ref{tab:system_params}, thereby maximizing the  receivable NLoS signal power and the overall spatial-temporal information gain at the \gls{UE}.  

%All the following simulations are reported using Monte Carlo method under random geometries (random \gls{UE}/\glspl{PT} locations and orientations) with $3$ \glspl{PT}.

\subsection{UE Position and Velocity Estimation Study}
All the following simulations are Monte Carlo simulations with $200$ independent channel realizations, including random positions of \gls{UE} and \glspl{PT}, to ensure statistically reliable estimates of PEB and VEB. The \gls{UE} and \glspl{PT} are generated to be within $[5,100]$m from the broadside of the BS.

We analyze the performance of \gls{UE} velocity estimation and investigate the inherent coupling between position and velocity across the \gls{MS} and \gls{BP} sensing modalities. Estimating velocity via \gls{MS} relies purely on the Doppler shift observed from the direct LoS echo, which only captures the radial component of the \gls{UE}'s motion vector. Consequently, if the \gls{UE} has a non-negligible tangential velocity, the monostatic FIM becomes ill-conditioned, leading to unbounded velocity error bounds. %We report results considering solving the beamformer in its full-dimensional format\footnote{A lower-complexity Codebook Power Allocation (CPA) can be also derived from (P1) where CPA restrict the transmit beamformer covariance to a fixed beam codebook and optimizes only the per-beam powers.}, as formulated in (P1). 
To evaluate this limitation and compare it against our proposed \gls{BP} framework, we consider two distinct velocity estimators\footnote{We note that while \gls{BP} estimates velocity instantaneously, both Snapshot \gls{MS} and Extended \gls{MS} inherently suffer from latency, as velocity can only be resolved after the second temporal observation.}:

\begin{itemize}

\item Snapshot \gls{MS}:  In this approach, we simplify  (P1) by completely removing velocity and Doppler from the estimation stage, effectively treating the \gls{UE} as static within each coherence interval. Two temporally separated \gls{MS} snapshots are then considered, each yielding a position estimate with its associated covariance. The velocity is subsequently inferred by finite differencing the two position estimates. 

\item Extended \gls{MS}:  This approach exploits Doppler measurements from two temporal instants under a constant-velocity model. Because the \gls{UE} moves between observations, the Doppler shifts project the velocity onto two distinct LoS directions, enabling the joint reconstruction of both radial and tangential velocity components.
\end{itemize}

%In Fig.~\ref{fig:2a}, and under a symmetric geometry, we show the radial VEB for both \gls{MS} and \gls{BP}. As $\alpha$ increases toward $1$, the radial VEB degrades for both approaches. This behavior stems from the fact that radial velocity estimation fundamentally relies on strong LoS energy. In \gls{MS}, the LoS path is the only source of Doppler information; thus, reducing \gls{MS} energy directly worsens radial velocity estimation accuracy. The FDB approach consistently achieves the lowest radial VEB. This is because FDB optimizes the transmit beamformer covariance continuously, enabling precise beam steering toward informative spatial directions. In contrast, CPA employs a fixed beam codebook and performs discrete power allocation across predefined beams.

%Fig.~\ref{fig:2b} illustrates the \gls{UE} PEB tradeoff between \gls{MS} and \gls{BP} for the original FDB problem and its CPA counterpart.
%It shows a conflicting objective structure resembling a Phillips-curve-type tradeoff: improving \gls{BP} performance necessarily degrades \gls{MS} performance and vice versa. \gls{BP} achieves the lowest range of PEB and VEB radial due to the geometric diversity from multiple paths and angles, which yields less correlated measurements and stronger Fisher information in both position and velocity domains.

Fig.\ref{fig:4}  illustrates the evolution of the PEB and VEB with respect to the number \glspl{PT}. In Fig.~3(a), the PEB for \gls{BP} is initially unresolved when a single \gls{PT} is present, since the system lacks sufficient geometric diversity to infer both position and synchronization parameters. Once $2$ or more targets are introduced, matching the minimum dimensional requirement, the \gls{BP} configuration rapidly becomes more informative and surpasses the \gls{MS} performance. In contrast, the PEB of \gls{MS} increases with the number of \glspl{PT}, as multiple spatial beams must be steered toward different directions, reducing the energy and directivity available along LoS path to the user equipment. 
The same trend is reflected in Fig.\ref{fig:4b} for the VEB. Both the snapshot and extended \gls{MS} estimators exhibit performance degradation with additional targets for the same reason: the transmit power and beamforming gain are increasingly distributed over non-LoS directions. The degradation is more pronounced for the extended \gls{MS} case, which strongly depends on the resolvability of the tangential velocity component, particularly sensitive to the \gls{UE}  relative position during the second observation interval. The snapshot \gls{MS}, in contrast, derives its velocity information indirectly from the position covariance, resulting in a more gradual performance loss as the number of \glspl{PT} grows.

Fig.~\ref{fig:5} analyzes velocity estimation. We plot the \gls{UE} VEB versus speed for Snapshot \gls{MS}, Extended \gls{MS}, and \gls{BP}. As the speed increases, the VEB degrades for Snapshot \gls{MS} and Extended \gls{MS}. This trend is explained by the reduced number of effective observation slots within a CPI as higher Doppler accelerates the phase rotation across symbols/subcarriers, which shortens the interval over which the assumed parametric model remains informative and thereby lowering the usable Fisher information. \gls{BP} consistently perform better than Snapshot \gls{MS} and Extended \gls{MS}.

%Fig.~\ref{fig:4}, we further examine the coupling between tangential and radial velocity versus speed. For \gls{BP}, this ratio remains approximately constant across speeds, indicating that the relative conditioning between tangential and radial components is largely geometry-driven and does not change with speed when both components are jointly supported by LoS and NLoS information. In contrast, for Extended \gls{MS}, the ratio decreases as speed increases, meaning that the tangential component becomes increasingly comparable to the radial one. This occurs because the \gls{UE} displacement between the two observation instants grows with speed, which increases the angular separation between the two LoS unit vectors and improves the geometric diversity of the two Doppler projections. 

\begin{comment}
\begin{figure}[!htb]
 \centering
  \hspace{-0.5cm}
 \label{fig:images2}
 \subfloat[]{%
      \includegraphics[width=0.25\textwidth]{figures/fig7a.pdf}}
      \label{fig1}
 \qquad
 \hspace{-0.9cm}
 \subfloat[]{%
      \includegraphics[width=0.243\textwidth]{figures/fig7b.pdf}}
      \label{fig3}

       \caption{Radial and Tangential Velocity}%
\end{figure}
\end{comment}

\subsection{\gls{MS}/\gls{BP} Allocation and Sequential Optimization}

In Fig.\ref{fig5}, we consider different UE velocities in $[5,30]$(m/s) and also different numbers of \glspl{PT} between $2$ and $4$, and we  compare the full \gls{MS}, full \gls{BP}, the shared sequential beamformer, and the separate sequential beamformer approaches by evaluating the \gls{UE} PEB, \gls{UE} VEB, \glspl{PT} PEB, and computation time. 
We can see that the shared and separate sequential beamformer approach consistently yields better performances compared to full \gls{MS} and full \glspl{BP}. In particular, both sequential designs achieve cm-level positioning accuracy for the \gls{UE} and the \glspl{PT} over the considered scenarios.
Parametrizing the \gls{BP} FIM via the separate and shared beamformers improves conditioning, enabling the \gls{UE} estimation of \glspl{PT} positions, unlike in the full \gls{BP} baseline.  Fig.~\ref{fig5} also reveals that full \gls{MS} performance degrades with decreasing $\rho_{\text{MS}}$ ratio, illustrating the fundamental tradeoff between monostatic prior quality and the posterior information available from the \gls{BP} stage in the sequential designs. The optimization problems (P2) and (P3) resolve this tradeoff by determining the optimal $\rho_{\text{MS}}^\star$ that minimizes the desired bounds.

Fig.~\ref{fig5a}, Fig.~\ref{fig5b}, and Fig.~\ref{fig5c} show that the shared sequential beamformer consistently outperforms the separate sequential design in both \gls{UE} and \glspl{PT} positioning, despite the latter having more spatial degrees of freedom.  This behavior originates from the joint spatio-temporal structure imposed by the shared beamformer, which maintains a consistent illumination pattern across the \gls{MS} and \gls{BP} stages. Because the same beams coherently interrogate the scene over time, the Doppler and delay shifts associated with both the \gls{UE} and \glspl{PT} become more accurately captured, effectively enhancing parameter resolvability. Furthermore, velocity estimation relies predominantly on the \gls{BP} Doppler measurement, to which the MS stage contributes only the radial component. By contrast, the separate approach commits $\mathbf{F}_{\text{MS}}^{\star}$ based solely on the MS cost, distributing power more broadly across directions and thus weakening the prior information passed to the \gls{BP} stage for each individual node.
Also, all the metrics show that  \gls{MS} prior dominates the lack of \gls{BP} symbols.

Finally, we compare the computational running time of the different designs in Fig.~\ref{fig5d}. The separate sequential beamformer exhibits a noticeably higher runtime than the shared counterpart due to it being implemented as a two-stage greedy optimization: it first optimizes $\mathbf{F}_{\text{MS}}$ and then, conditioned on this choice, optimizes $\mathbf{F}_{\text{BP}}$ over a second search space. In contrast, the shared formulation solves a single global optimization problem over a common beamformer used in both stages, which effectively reduces the dimensionality and decouples the search from stage-specific reconfiguration.

\section{Conclusion}
In this paper, we proposed a velocity-aware, sequential beamforming framework for joint monostatic sensing (MS) and bistatic positioning (BP). By executing MS first, the base station constructs a spatial-temporal prior regarding the environment and the mobile user, which is then utilized to regularize the subsequent BP estimation. Our results demonstrate that optimizing a single shared beamformer globally across both stages outperforms stage-specific greedy designs. This shared approach effectively maximizes the joint posterior information, achieving centimeter-level positioning and robust velocity estimation with reduced computational complexity. 
Future work will extend this framework to a dynamic time-series model, continuously adapting symbol allocation to the evolving network geometry, ultimately evolving into a multi-static online beamforming. %architecture where real-time MS measurements act as an iteratively updated prior, coherently accumulating knowledge to dynamically guide network-wide sensing.

\section*{Acknowledgment}
This work is supported by the EU-HORIZON-MSCA-2022-DN 6thSense project, Grant Agreement N°101119652, under the EU's Programme for Research and Innovation.

\balance
\bibliographystyle{IEEEtran}
\bibliography{ref}

\end{document}